\documentstyle[epsfig]{article}
\begin{document}
\begin{center}
{\large\bf Quantum Critical Behavior of Two Coupled
 Bose-Einstein Condensates}
\vskip .6cm {\normalsize Feng Pan,$^{1,2}$ and J. P. Draayer$^{2}$ }
\vskip .2cm {\small
$^{1}$Department of Physics, Liaoning Normal University, Dalian
116029, P. R. China\vskip .1cm
$^{2}$Department of Physics and Astronomy,
Louisiana State University, Baton Rouge, LA 70803-4001}
\end{center}
\vskip .5cm
{\bf Abstract:~}\normalsize
The quantum critical behavior of the Bose-Hubbard model for
a description of two coupled Bose-Einstein condensates is
studied within the framework of an algebraic theory. Energy
levels, wavefunction overlaps with those of the Rabi and Fock
regimes, and the entanglement are calculated exactly as functions
of the phase parameter and the number of bosons. The results
show that the system goes though a phase transition and that
the critical behavior is enhanced in the thermodynamic limit.

\vskip .5cm

As is well-known, the two-site Bose-Hubbard model$^{[1]}$
can be used to describe pair tunneling between two
superconductors through an insulating junction,
trapped ultra-cold bosonic gases, etc.
Furthermore, it can also be used to prepare
macroscopically entangled states.$^{[2]}$
The two-site Bose-Hubbard model has been
investigated widely by many authors using various methods, such as the
Gross-Pitaevskii approximation,$^{[3]}$ mean-field theory,$^{[4-5]}$
the quantum phase model,$^{[6]}$ and the Bethe ansatz method.$^{[7]}$
In [8], the temporal evolution of the expectation value for the relative
number of particles between the two condensates for different choices
of the coupling parameter and distinct initial states is
analyzed. Also, quantum phase transitions that
occur at zero temperature as a function of a coupling constant
have become important in connection with various
quantum many-body systems, such as the quantum Ising and
rotor models,$^{[9]}$ Fermi liquids,$^{[10]}$, and atomic nuclei.$^{[11]}$
There are distinct features in these systems at
the critical points. The main purpose of
the present paper is to study the critical behavior
of the two-site Bose-Hubbard model as a function of
the coupling parameter and the total number of
bosons.

Specifically, we consider the two-site Bose-Hubbard Hamiltonian with

$$\hat{{\cal
H}}=-E_{J}(c^{\dagger}d+d^{\dagger}c)+E_{c}(c^{\dagger}cc^{\dagger}c+
d^{\dagger}dd^{\dagger}d),\eqno(1)$$
where $c^{\dagger}$ ($c$) and $d^{\dagger}$ ($d$) are boson creation
(annihilation)
operators in two traps or different hyperfine states. $E_{J}$ is
related to the Josephson coupling exchanging bosons between the
two states, and $E_{c}$ is related to the charging energy.
In the present work we focus on the case where the effective
interaction energy for the internal Josephson dynamics is
negative,$^{[12]}$ $E_c < 0$.
In order to study the transitional patterns of the system, the
Hamiltonian (1) is re-parameterized as

$$\hat{H}=\hat{{\cal
H}}/E_{0}=-(1-x)(c^{\dagger}d+d^{\dagger}c)-{4x\over{n+1}}(c^{\dagger}cc^{\dagger}c+
d^{\dagger}dd^{\dagger}d),\eqno(2)$$
where $E_0$ is a constant in arbitrary unit, $n$ is the total
number of bosons, and $0\leq x\leq 1$ is the phase parameter.
The division of the second term in (2) by $n+1$ serves to ensure that
the boson rank
of the two terms in $\hat{H}$ is the same, thereby making comparisons
of results as
a function of boson number more meaningful.
For large values of $n$, this system is in the Rabi regime when $x\sim 0$,
the Fock regime when $x\sim 1$, and the Josephson regime when
$0< x< 1$.

When $x=0$, the system is
in the Rabi regime with eigenstates given by

$$\vert x=0; n_1,n_2\rangle=
{1\over{2^{n/2}}}(c^\dagger+d^\dagger)^{n_1}
(c^\dagger-d^\dagger)^{n_2}\vert 0\rangle,\eqno(3)$$
where $n=n_1+n_2$, and $\vert 0\rangle$ is the boson vacuum state
which is never degenerate. The corresponding eigenenergy of (2) is

$$E_{n_1,n_2}(x=0)=n_2-n_1.\eqno(4)$$

When $x=1$, the system is
in the Fock regime with eigenstates given by

$$\vert x=1;n_1,n_2\rangle={c^{\dagger n_{1}}d^{\dagger n_{2}}
\over{\sqrt{n_1!n_2!}}}\vert 0\rangle,\eqno(5)$$
which is two-fold degenerate under the permutation $n_1\rightleftharpoons n_2$.
The corresponding eigenenergy is

$$E_{n_1,n_2}(x=1)=-{4\over{n+1}}\left(n_1^2+n_2^2\right).\eqno(6)$$

In the Josephson regime, $0< x< 1$, the unitary transformation [13]

$$c={1\over{\sqrt{2}}}(a-ib),~
d={1\over{\sqrt{2}}}(a+ib)\eqno(7)$$
can be used to rewrite Hamiltonian (2) as

$$\hat{H}=(1-x)(b^\dagger b-a^\dagger
a)+{2x\over{n+1}}S^{+}(0)S^{-}(0)-{4n^{2}\over{n+1}}x,\eqno(8)$$
where

$$S^{+}(0)=b^{\dagger 2}+a^{\dagger
2},~S^{-}(0)=b^{2}+a^{2}\eqno(9)$$
are boson pairing operators.
In a manner similar to what was done in [7] and [14], it can be shown that the
Bethe ansatz eigenvectors used for diagonalizing the Hamiltonian (8)
may be written as

$$\vert x; \zeta; n=2k+\nu_1+\nu_2
\rangle={\cal N}S^{+}(y^{(\zeta)}_{1})S^{+}(y^{(\zeta)}_{2})\cdots
S^{+}(y^{(\zeta)}_{k})\vert \nu_1,\nu_2\rangle,\eqno(10)$$
where ${\cal N}$ is the normalization constant,
$\vert \nu_1,\nu_2\rangle=a^{\dagger\nu_1}b^{\dagger\nu_2}\vert 0\rangle$
with $\nu_{i}=0$ or $1$ for $i=1, 2$
is the boson pairing
vacuum state that satisfies

$$a^{2}\vert \nu_1,\nu_2\rangle=b^{2}\vert \nu_1,\nu_2\rangle=0,\eqno(11)$$
and

$$S^{+}(y^{(\zeta)}_{i})={b^{\dagger 2}\over{1-y^{(\zeta)}_{i}}}+{a^{\dagger
2}\over{1+
y^{(\zeta)}_{i}}},\eqno(12)$$
in which $y^{(\zeta)}_{i}$ ($i=1,2,\cdots,k$)
are spectral parameters that are to be determined,
and $\zeta$ is an additional quantum number for distinguishing
different eigenvectors with the same quantum number $k$.
It can then be verified by using the corresponding
eigen-equation that (10) is the solution when the
spectral parameters $y_{i}^{(\zeta)}$ ($i=1,2,\cdots, k$)
satisfy the following set of equations:

$${2x\over{n+1}}\left( {2\nu_{2}+1\over{1-y_{i}^{(\zeta)}}}+
  {2\nu_{1}+1\over{1+y_{i}^{(\zeta)}}}\right)={1-x\over{y_{i}^{(\zeta)}}}
  +8x\sum_{j(\neq i)}{y_{j}^{(\zeta)}
  \over{y_{i}^{(\zeta)}-y_{j}^{(\zeta)}}}\eqno(13)$$
for $i=1,2,\cdots,k$ with the corresponding eigen-energy given by

$$E^{(\zeta)}_{k}(x)=(1-x)\sum_{i=1}^{k}{2\over{y_{i}^{(\zeta)}}}+
(1-x)(\nu_2-\nu_1)-{4n^{2}\over{n+1}}x.\eqno(14)$$
It should be noted that the solutions (13) and (14) are not
valid in  the $x=0$ (Rabi) or $1$ (Fock) limits.

   To explore transitional patterns, some low-lying
energy levels of the system for $x\in [0,1]$
with $n=10$, $40$, $100$, and $160$, respectively,
were calculated.  The results, which are shown
in Fig. 1, clearly show that
there is a minimum in the excitation energies
around $x\sim 0.25-0.35$ which corresponds to the
Josephson critical region.
While for small boson numbers the critical region
is not very well defined, it becomes clearer and sharper
with increasing $n$ values.  Also, the energy level
density in the critical region increases
with increasing $n$.

To probe the nature of the critical point behavior more deeply,
overlaps $\vert\langle x; n\vert x_{0};n\rangle\vert$
for the ground states with $x_0 = 0$ or $1$ for
$n=10$, $40$, $80$, and $120$, respectively, were calculated.
The results, which are given in Fig. 2, show that there is a
crossover point at a certain nonzero amplitude
for the overlaps $\vert\langle x; n\vert x_{0}=0;n\rangle\vert$
and $\vert\langle x; n\vert x_{0}=1;n\rangle\vert$ when
$n$ is relatively small, which yields to a cross-over region
with near zero amplitude when $n$ becomes large.
This critical region in the large $n$ limit
should be regarded as a two phase (Rabi and Fock)
coexistence region in analogue to the situation
occurring in other finite boson systems.$^{[11, 15]}$
Furthermore, there is a sharp change in
$\vert\langle x; n\vert x_{0}=0;n\rangle\vert$
around a critical point $x_{c}\sim 0.25$
for large $n$. These results suggest that the largest
absolute value of the derivative of
$\vert\langle x; n\vert x_{0}=0;n\rangle\vert$
with respect to $x$ occurs around the critical
point in the thermodynamic limit.

As is known, the entanglement measure for any pure bipartite system
is defined by

$$\eta = -{\rm Tr}(\rho_{c} \log_{N}\rho_{c} ) =-{\rm Tr}(\rho_{d}
\log_{N}\rho_{d}),\eqno(15)$$
where $\rho_{c}$ is the reduced density matrix obtained by taking the
partial trace over
the subsystem $d$, and similarly for $\rho_{d}$, and $N$ is the
total number of the Fock states for given $n$. We use the
logarithm to the base $N$ instead of base $2$ to ensure that
the maximal measure is normalized to $1$.
Fig. 3 shows the entanglement measure of the system
as a function of $x$ for $n=10$, $40$, $80$, and $120$,
respectively. It is obvious that
there is always a peak in the measure at or near
the critical point, which is consistent with
the so-called critical point entanglement.$^{[16]}$
For small $n$, the maximal value of the measure
is near $1$, while it decreases with increasing values of
$n$. The peak grows sharper with increasing values of $n$, tracking
the behavior of
the overlap $\vert\langle x; n\vert x_{0}=0;n\rangle\vert$
and the excitation energies around the critical point.

In summary, we have studied the quantum critical behavior of
the two-site Bose-Hubbard model
within the framework of an algebraic theory.
Energy levels, overlaps of the wavefunctions with those in both the Rabi
and Fock regimes, and the entanglement were calculated exactly as a function
of the phase parameter and the total number of bosons.
The results show not only
the quantum phase transition patterns
of the model, but also that
the critical behavior
are greatly enhanced in the thermodynamic (large $n$) limit.
This enhancement of critical phenomena with increasing
of $n$ should be common in other
finite quantum many-body systems, as recently shown
in the interacting boson model for atomic nuclei.$^{[17]}$
More importantly, the results suggest that while the quantum phase
transition of the two-site Bose-Hubbard model
between the Rabi and the Fock regimes is rather smooth
for small values of $n$, it becomes clearer and sharp
in the thermodynamic limit. Therefore, such
quantum phase transition should be observed macroscopically.

\vskip .2cm
Support from the U.S. National Science Foundation
(0140300), the Southeastern Universitites Research
Association, the Natural Science Foundation of China
(10175031), the Natural Science Foundation of Liaoning
Province (2001101053), the Education Department
of Liaoning Province (202122024), and the LSU--LNNU joint
research program (C164063) is acknowledged.

\vskip .5cm

\def\HT{\bf\relax}
\def\REF#1{\small\par\hangindent\parindent\indent\llap{#1\enspace}\ignorespaces}

\section*{Reference}

\noindent\REF{[1]} M. P. A. Fisher, P. B. Weichman, G. Grinstein,
and D. S. Fisher, Phys. Rev. B {\bf 40} (1989) 546.

\REF{[2]}L. You, Phys. Rev. Lett. {\bf 90} (2003) 030402.

\REF{[3]} A. J. Leggett, Rev. Mod. Phys. {\bf 73} (2001) 307.

\REF{[4]} G. J. Milburn, J. Corney, E. M. Wright and D. F. Walls,
Phys. Rev. A {\bf 55}
(1997) 4318.

\REF{[5]} A. P. Hines, R. H. McKenzie and G. J. Milburn, Phys. Rev. A
{\bf 67} (2003)
013609.
\REF{[6]}  J. R. Anglin, P. Drummond, A. Smerzi, Phys. Rev. A {\bf 64} (2001)
063605.

\REF{[7]} H.-Q. Zhou, J. Links, R. H. McKenzie and X. -W. Guan, J.
Phys. A: Math. Gen. {\bf
36} (2003) L113.

\REF{[8]} A. P. Tonel, J. Links, and A. Foerster,
arXiv:quant-ph/0408161.

\REF{[9]}S. Sachdev, {\it Quantum Phase Transitions} (Cambridge
Univ. Press, Cambridge, 1999).

\REF{[10]} J. A. Hertz, Phys. Rev. B {\bf 14} (1976) 1165.

\REF{[11]} F. Iachello and N. V. Zamfir, Phys. Rev. Lett. {\bf 92}
(2004) 212501.

\REF{[12]} S. Kohler and F. Sols, Phys. Rev. Lett. {\bf 89} (2002) 060403.

\REF{[13]} G. Ortiz, R. Somma, J. Dukelsky, and S. Rombouts,
arXiv:cond-mat/0407429.

\REF{[14]} Feng Pan and J. P. Draayer, Nucl. Phys. A {\bf  636}
(1998) 156.

\REF{[15]} J. Jolie, P. Cejnar, R. F. Casten, S. Heinze,  A. Linnemann, and
V.Werner, Phys. Rev. Lett. {\bf 89} (2002) 182502.

\REF{[16]} I. Bose and E. Chattopadhyay,  Phys. Rev. A {\bf 66} (2002) 062320.

\REF{[17]} D. J. Rowe, Phys. Rev. Lett. {\bf 93} (2004) 122502.

\newpage
\begin{center}
\includegraphics[totalheight=9cm,width=9.cm]{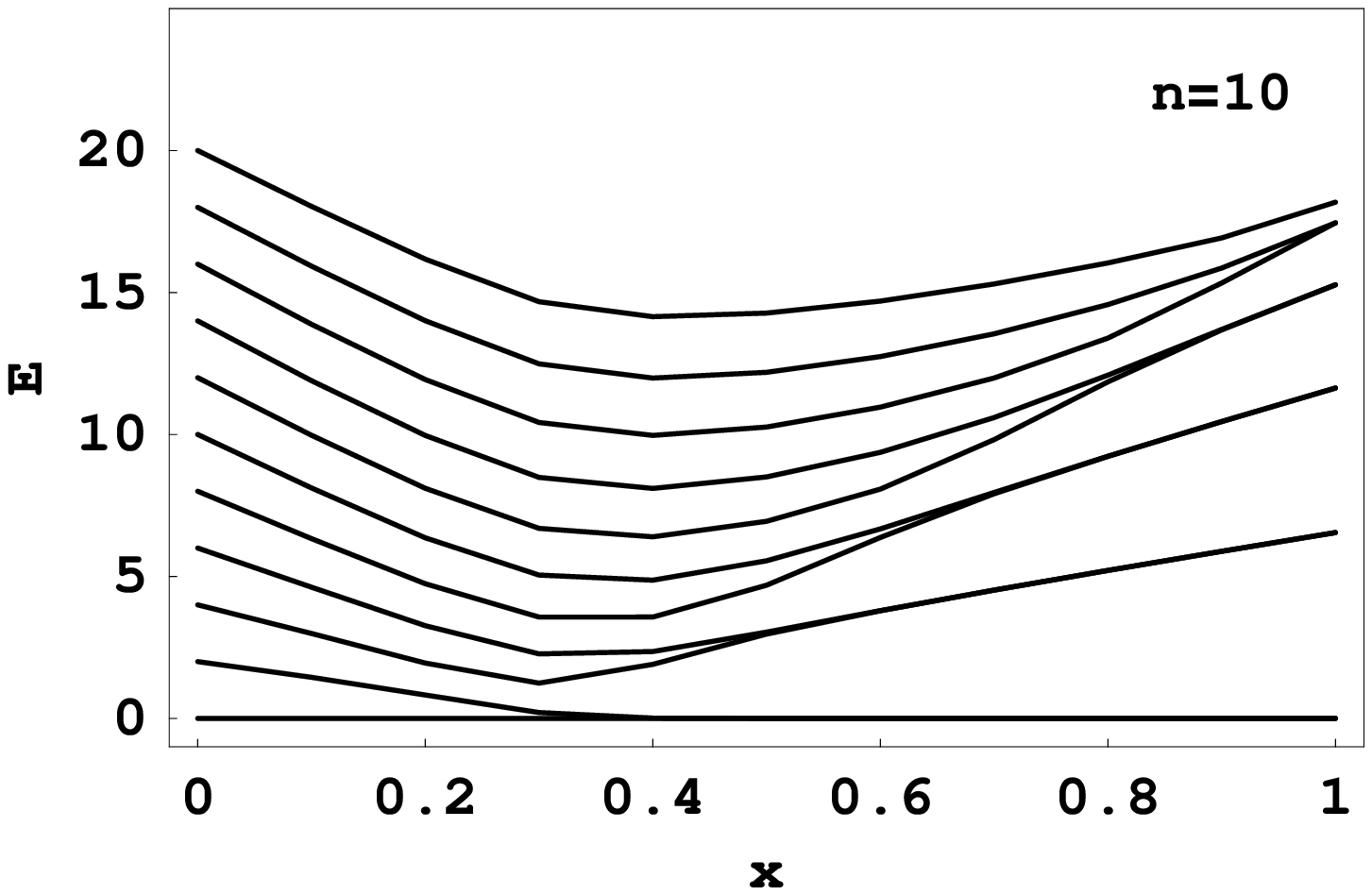}\includegraphics[totalheight=9.3cm,width=9.cm]{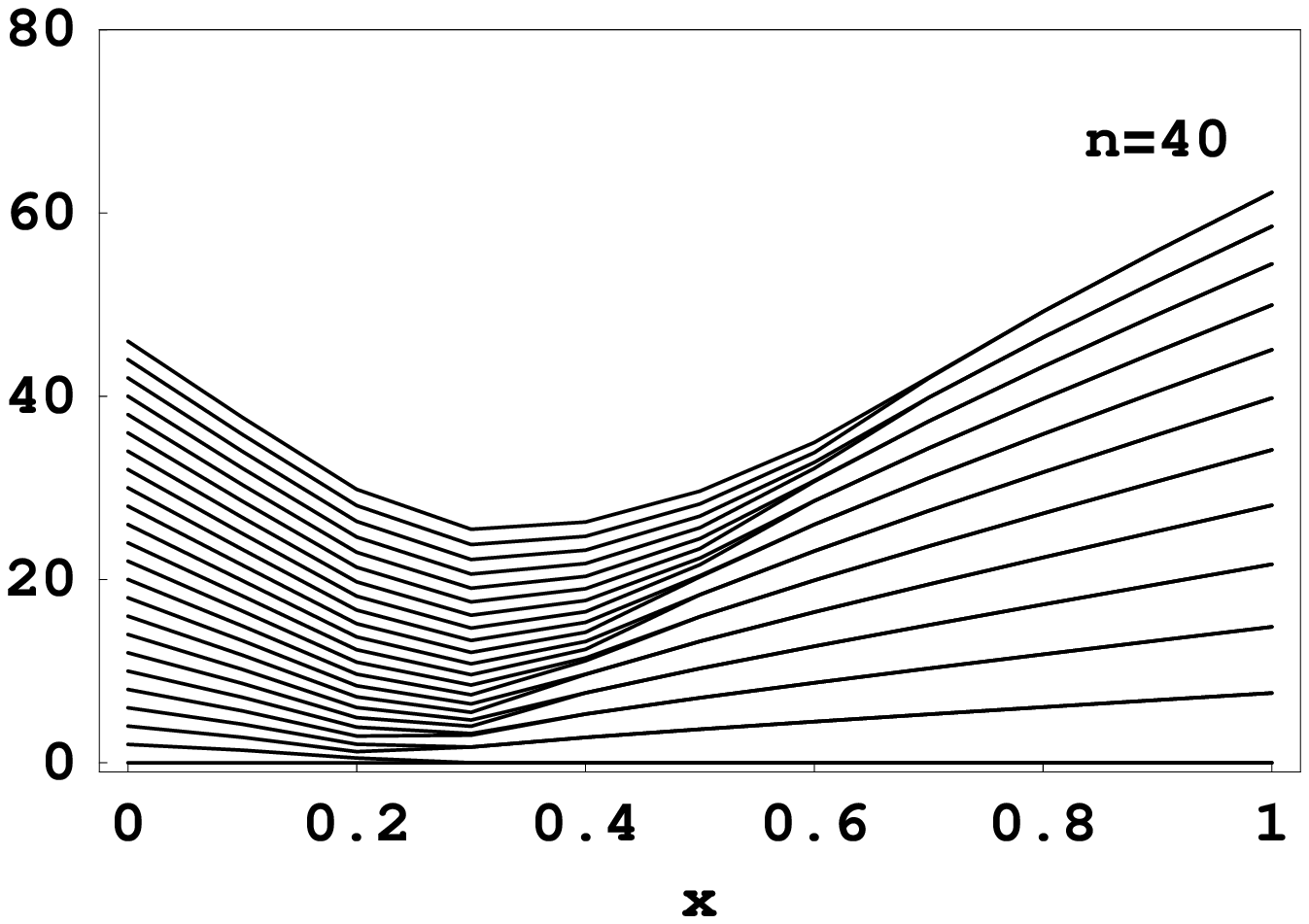}\\
\includegraphics[totalheight=10cm,width=9.cm]{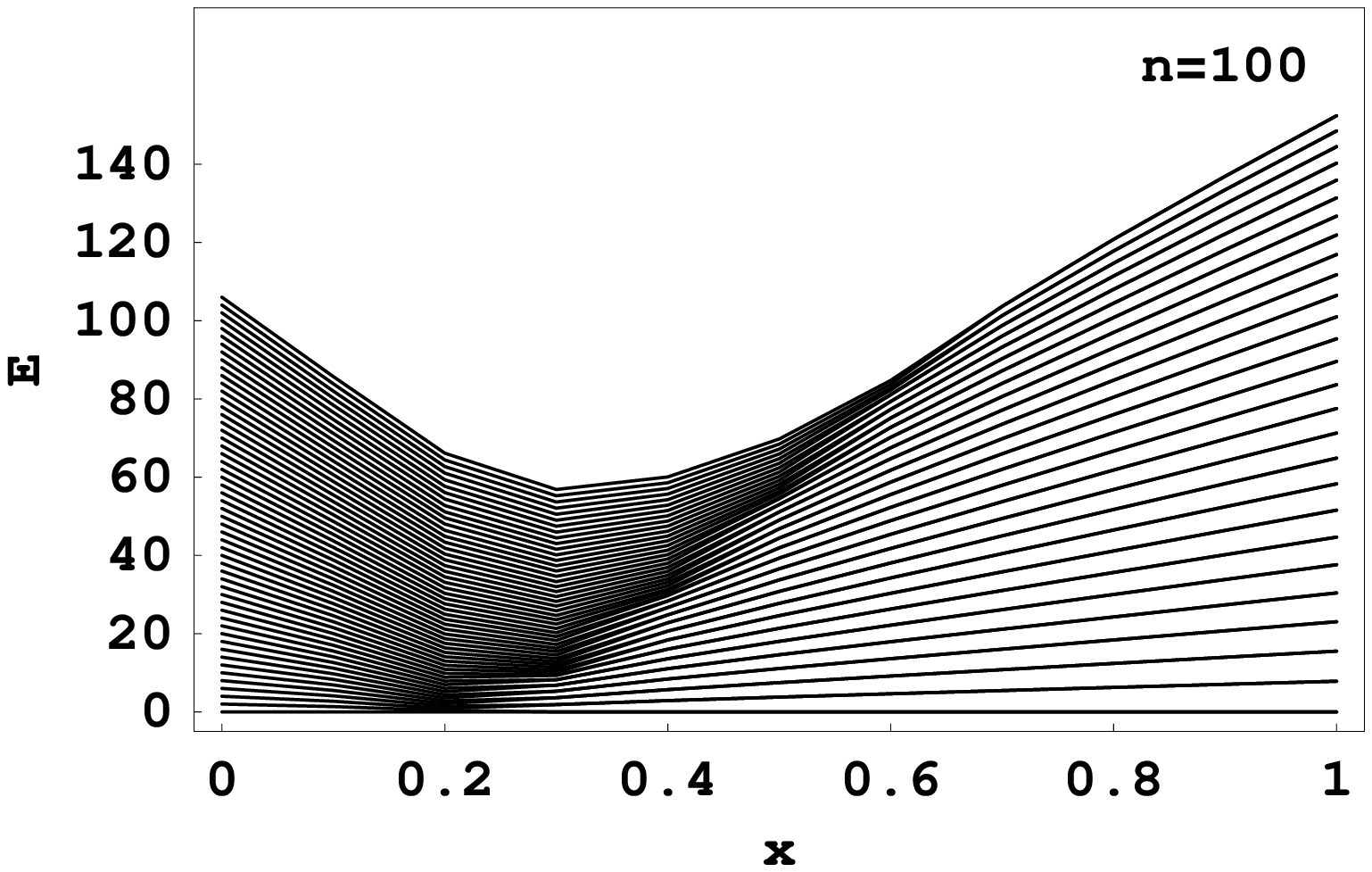}\includegraphics[totalheight=10cm,width=9.cm]{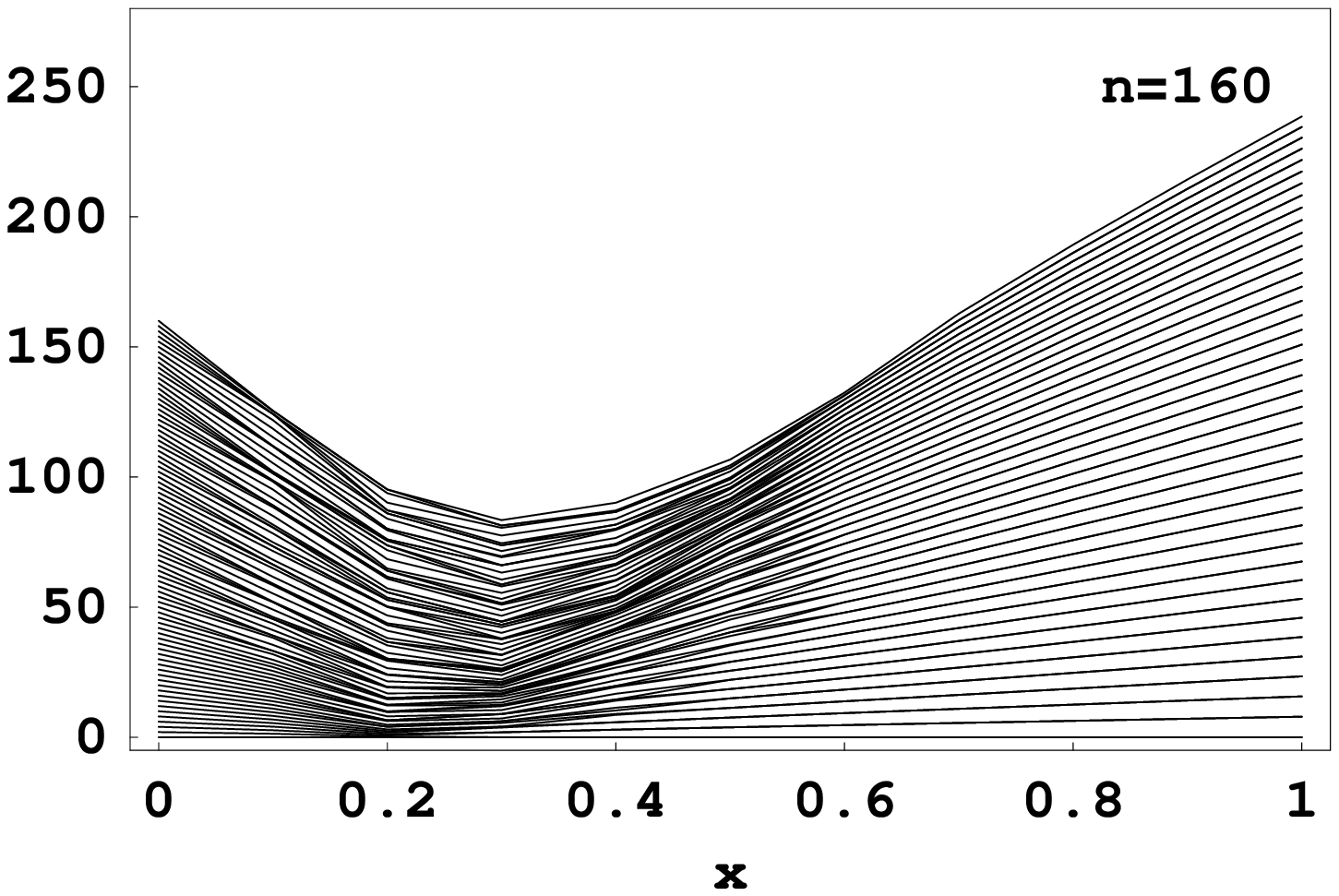}\\
\end{center}
{\bf Fig. 1}. {\small Some low-lying energy levels of  the two-site
Bose-Hubbard model
with the Hamiltonian given by (2) as functions of $x$  for $n=10$,
$40$, $100$, and $160$, respectively. }

\begin{center}
\includegraphics[totalheight=5cm,width=7cm]{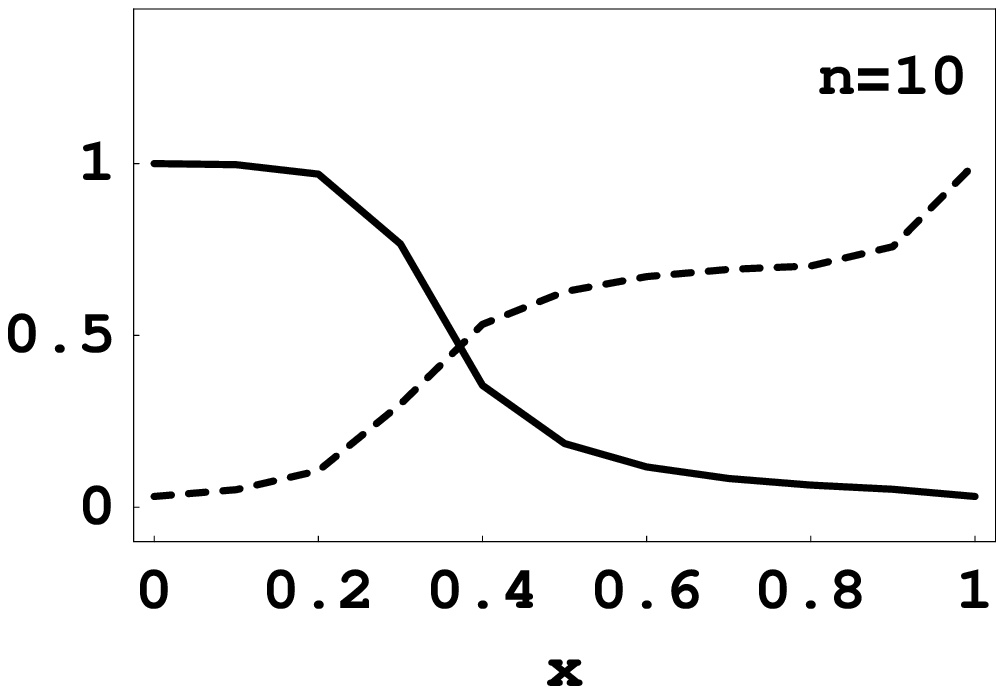}\includegraphics[totalheight=5cm,width=7.cm]{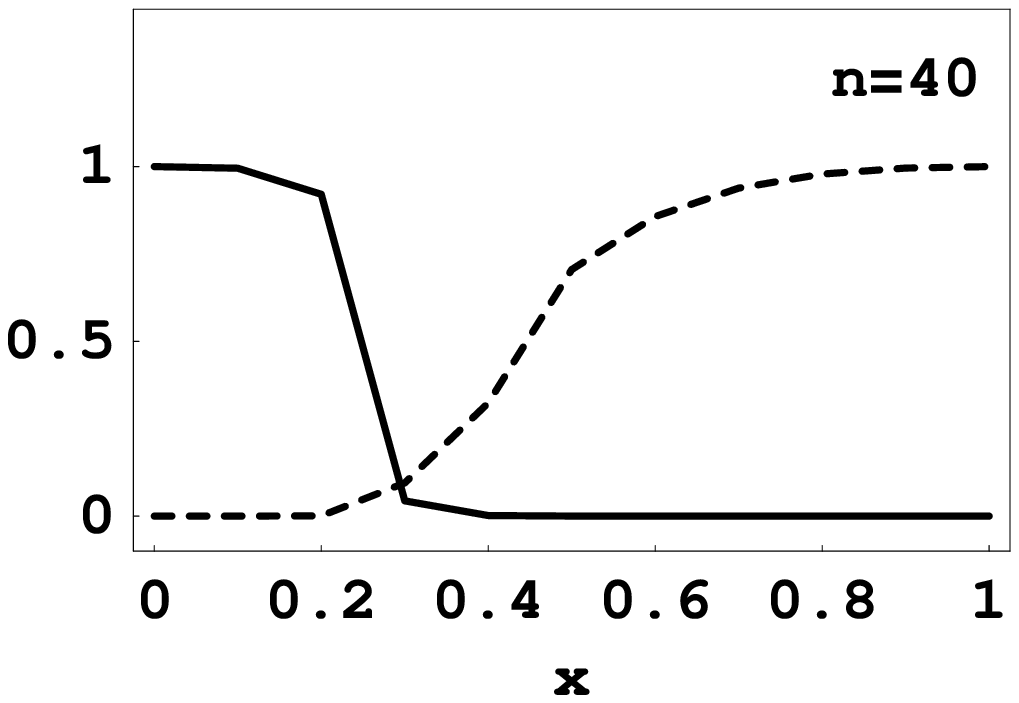}\\
\includegraphics[totalheight=5cm,width=7cm]{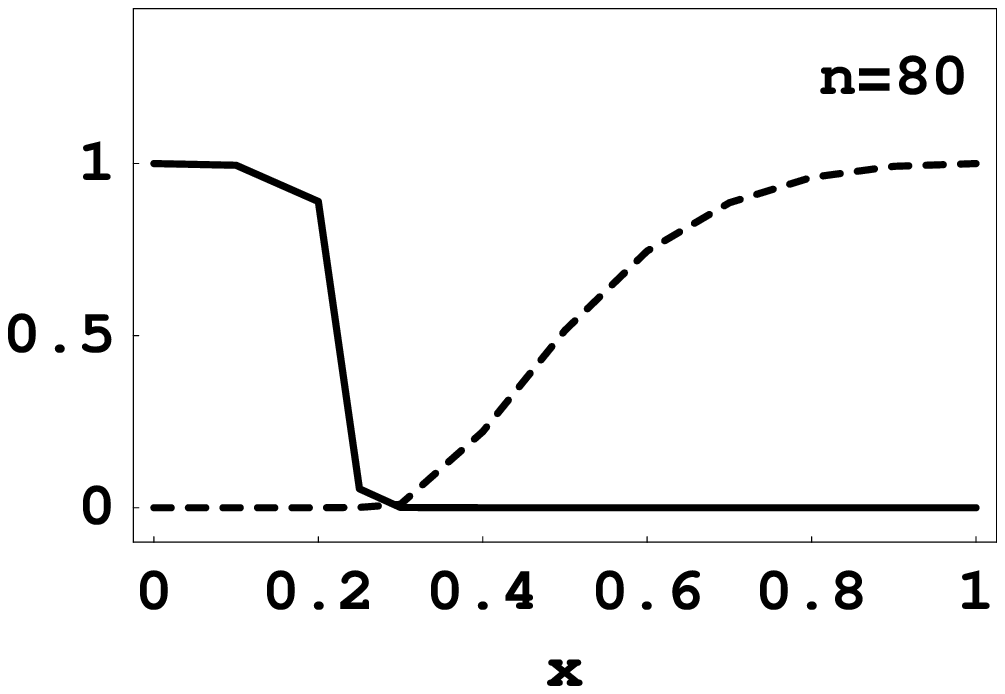}\includegraphics[totalheight=5cm,width=7.cm]{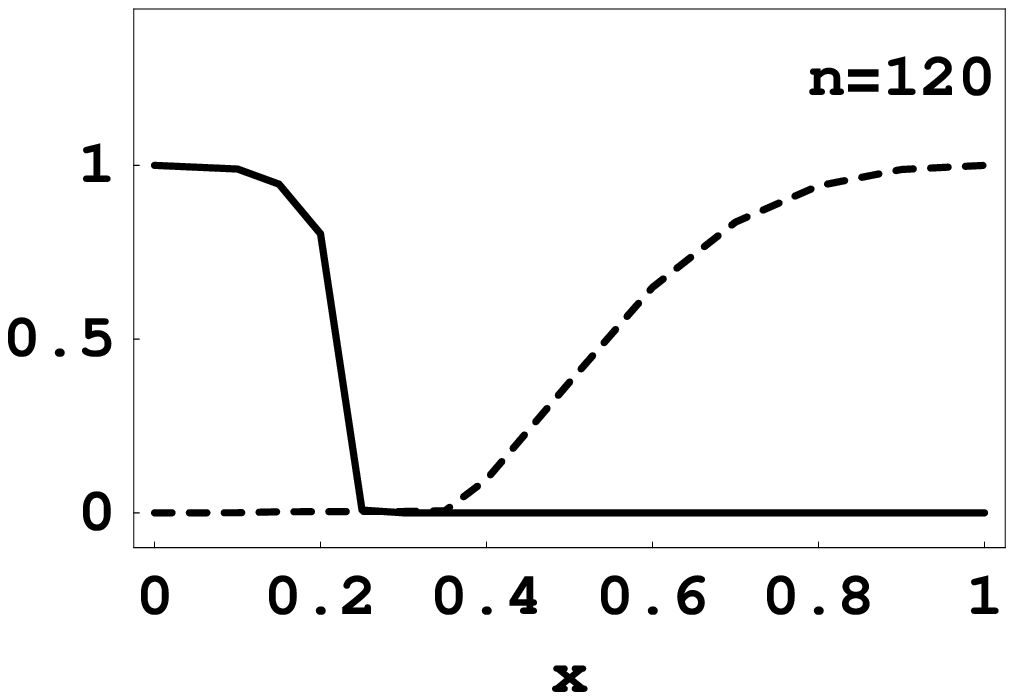}\\
\end{center}
{\bf Fig. 2}. {\small  Overlaps $\vert\langle x; n\vert x_{0};n\rangle\vert$
for the ground states with $x_0 = 0$ or $1$ for $n=10$, $40$, $80$, and
$120$, respectively.  The full lines represent the curves of
$\vert\langle x; n\vert x_{0}=0;n\rangle\vert$,
and the dotted lines represent those of $\vert\langle x; n\vert
x_{0}=1;n\rangle\vert$. }

\begin{center}
\includegraphics[totalheight=5cm,width=7cm]{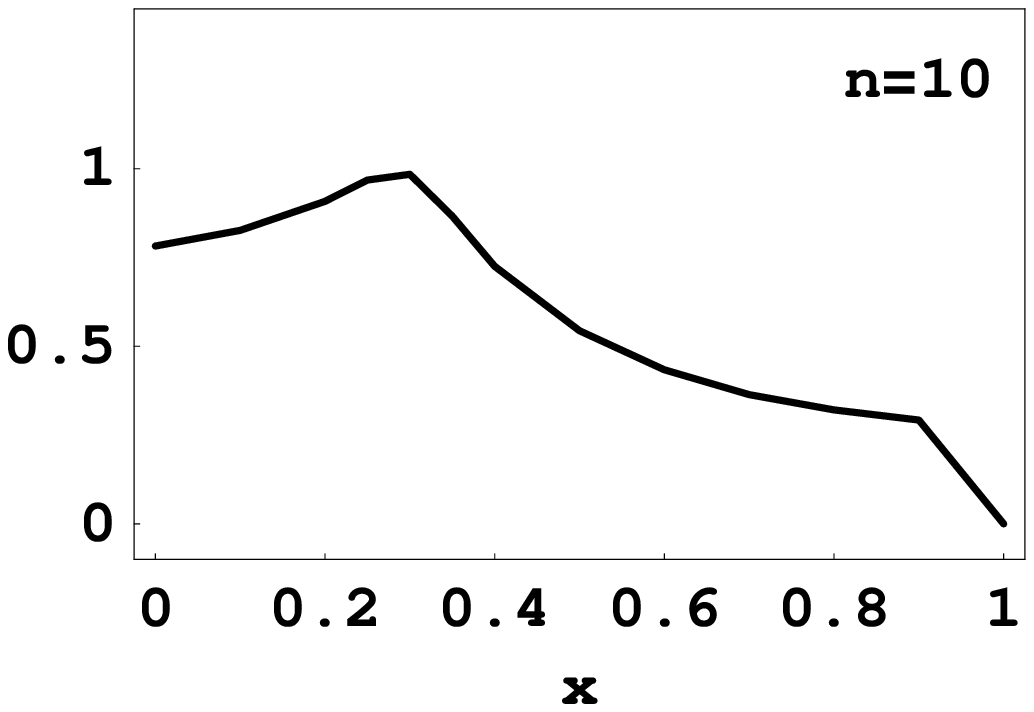}\includegraphics[totalheight=5cm,width=7.cm]{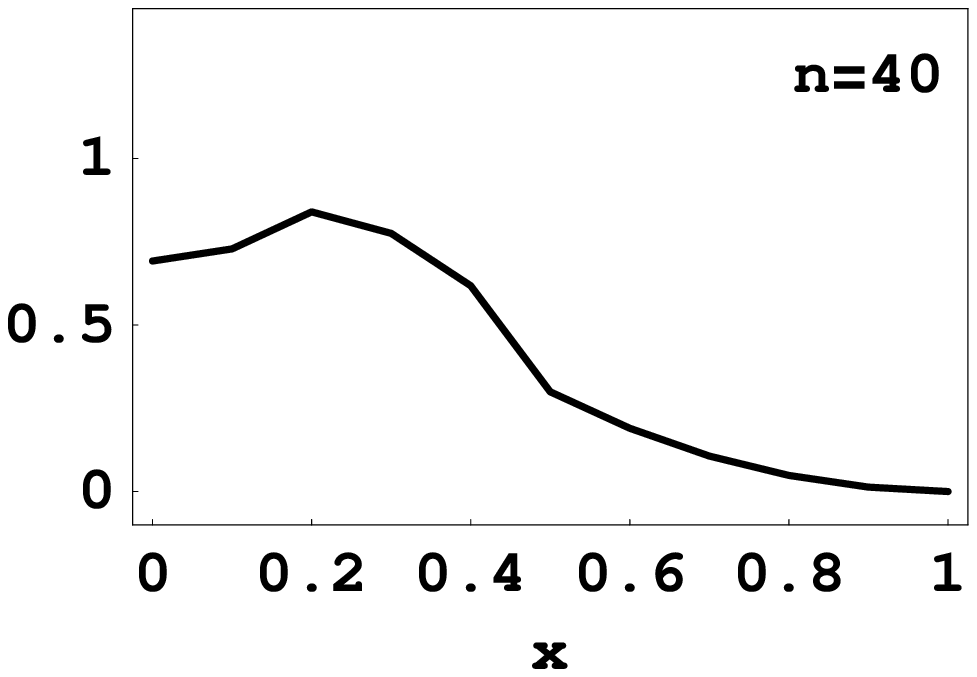}\\
\includegraphics[totalheight=5cm,width=7cm]{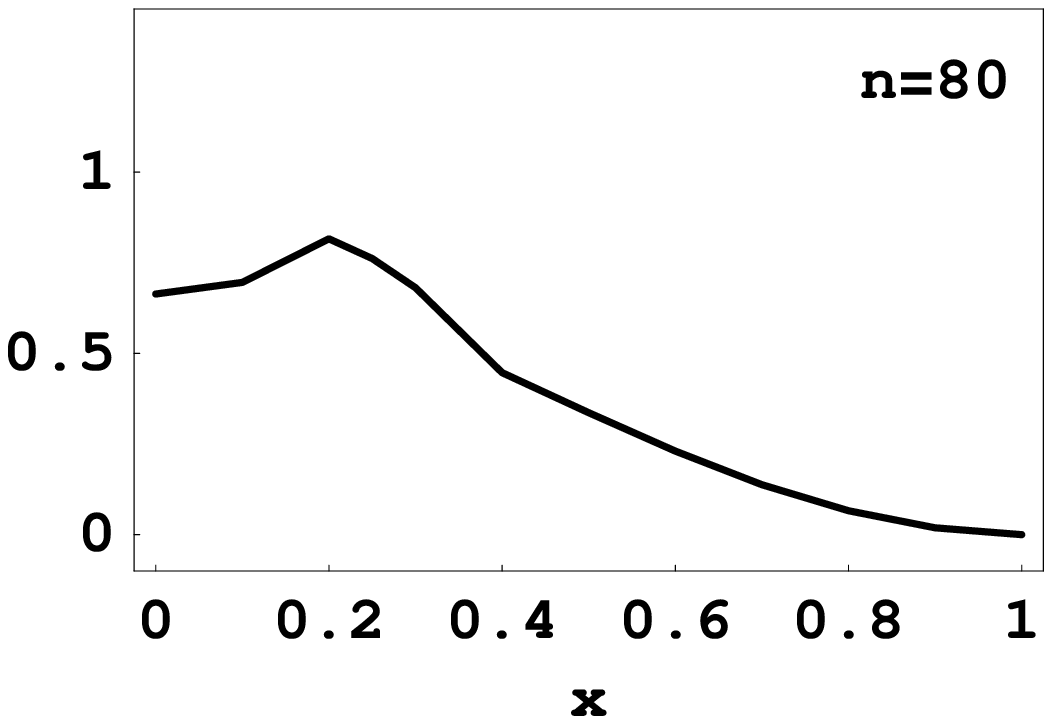}\includegraphics[totalheight=5cm,width=7.cm]{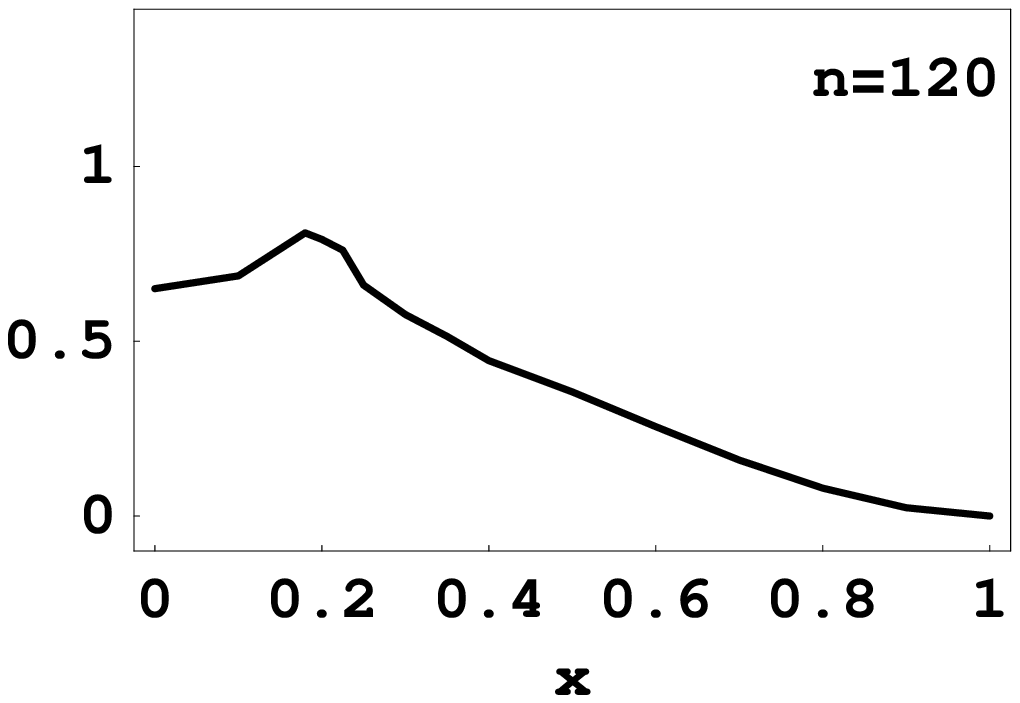}\\
\end{center}
{\bf Fig. 3}. {\small The entanglement measures of the model as a
function of $x$ for $n=10$, $40$,
$80$, and $120$, respectively.}

\end{document}